\documentclass[10pt,tightenlines,eqsecnum,floats,aps,amsmath,amssymb,nofootinbib,prd,showpacs]{revtex4}
\usepackage{amssymb}
\usepackage{stmaryrd}
\usepackage{amsmath}
\usepackage{amsfonts}
\usepackage{mathrsfs}

\usepackage{amsmath,amssymb,amsfonts}
\usepackage{graphicx}

\def\be{\begin{equation}}
\def\ee{\end{equation}}
\def\ba{\begin{eqnarray}}
\def\ea{\end{eqnarray}}

\def\z{\zeta}
\def\Hkw{\mathcal{H}^{\rm WDW}_{\rm kin}} 
\def\Hkl{\mathcal{H}^{\rm LQC}_{\rm kin}} 
\def\a{\alpha}
\def\c{\sigma}
\def\d{\delta}
\def\z{\zeta}
\def\e{\epsilon}
\def\TW{\widehat{\Theta}_{\rm WDW}}
\def\TL{\widehat{\Theta}_{\rm LQC}}
\def\CW{\widehat{C}_{\rm WDW}}
\def\CL{\widehat{C}_{\rm LQC}}
\def\D{\Delta}

\newcommand{\ket}[1]{\vert{#1} \rangle} 
\newcommand{\bra}[1]{\langle{#1}\,\vert} 
\newcommand{\obra}[1]{({#1}\vert} 
\newcommand{\inner}[2]{{\langle {#1} \vert  {#2} \rangle}} 
\newcommand{\opelem}[3]{{\langle {#1} \vert  {#2}  \vert  {#3} \rangle}} 
\newcommand{\norm}[1]{\|{#1}\|}
\newcommand{\com}[1]{\Psi_{z_{#1}}}
\newcommand{\cog}[1]{\underline{\Psi}_{\zeta_{#1}}}
\newcommand{\cow}[1]{\Psi_{\eta_{#1}}}

\newcommand{\eqnref}[1]{Eq.~(\ref{#1})}

\begin{document}

\title{Coherent State Functional Integrals in Quantum Cosmology}

\author{Li Qin}\email{qinli051@163.com}
\author{Yongge Ma\footnote{Corresponding author}}\email{mayg@bnu.edu.cn}
\affiliation{Department of Physics, Beijing Normal University, Beijing 100875, China}

\begin{abstract}
Coherent state functional integrals for the minisuperspace models of
quantum cosmology are studied. By the well-established canonical
theories, the transition amplitudes in the path-integral
representations of Wheeler-DeWitt quantum cosmology and loop quantum
cosmology can be formulated through group averaging. The effective
action and Hamiltonian with higher-order quantum corrections are
thus obtained in both models within the scheme of Gaussian coherent states. It turns out that for a non-symmetric
Hamiltonian constraint operator, the \emph{Moyal (star)-product} emerges naturally in the
effective Hamiltonian. This reveals the intrinsic relation among
coherent state functional integral, effective theory and
\emph{star-product}. Moreover, both the resulted effective theories imply a
possible quantum cosmological effect in large scale limit under certain condition.


\end{abstract}

\maketitle

\section{Introduction}
One of the most fundamental issues in modern physics is quantum
gravity. Among various approaches to quantum gravity, the viewpoint
of nonperturbative quantization has been received increased
attention. The first significant attempt of this kind was presented in 1960s \cite{DeWitt}.
Based on the ADM Hamiltonian formalism of general
relativity (GR) and the Dirac's generalized Hamiltonian quantization
theory, which is in principle applicable to constrained system,
Wheeler and DeWitt proposed a wave functional description of
gravitational field. In this quantum geometrodynamical approach,
the whole universe is described by a wave function which is defined on superspace
- the space of all 3-metrics
and matter field configurations. The dynamics is encoded in the
second-order differential Wheeler-DeWitt (WDW) equation. Despite the
elegant form of WDW theory, it encountered a number of fundamental
problems, such as the physical meaning of quantum constraint
equation, ordering of operators and problem of time
\cite{WDW-Review}. To overcome these obstacles, especially to impose
boundary conditions on WDW equation to find solutions, there were
some attempts appealing to the path integral formalism. Hartle and
Hawking suggested an Euclidean path integral representation
\cite{Hawking}, which gave the ``no-boundary" proposal for the wave
function. However, this approach can not be evaluated exactly in
general cases as the integral was usually badly divergent. To
alleviate the difficulty, Halliwell derived a Lorenzian
path integral formalism for minisuperspace models \cite{Halliwell},
which revealed the relationship between the choice of measure in the
path integral and the operator ordering in the WDW equation.

In the last two decades, an alternative background independent
approach developing rapidly is loop quantum gravity (LQG)
\cite{Rovelli,A-L,Thiemann,Han}. The starting point of LQG is the
Hamiltonian \emph{connection dynamics} of GR rather than the ADM
formalism. In this framework, GR looks like a gauge field theory with $SU(2)$ as its internal gauge group.
By taking the holonomy of $su(2)$-connection $A_{a}^{i}$
and flux of densitized triad $E^{b}_{j}$ as basic variables, the
quantum kinematical framework of LQG has been rigorously
constructed, and the Hamiltonian constraint operator can also be
well defined to represent quantum dynamics. Moreover, a few
physically significant results, especially the
\emph{resolution of big bang singularity}, have been obtained in the
minisuperspace models of loop quantum cosmology (LQC)
\cite{Boj2,aps1}. The quantum bounce replacement of big bang and its properties are being studied from different prescriptions of LQC \cite{Yang,k-l-p,vara,semilattice,scatt-lqc,lqc-comparative}. 
Effective equations were also derived in isotropic models
\cite{effective1,effective2,Ding}, which
predict evolution of universe with quantum corrections and
shed new lights on the singularity resolution. We refer to \cite{Boj,AS} for overviews on the recent progress in LQC.
Besides the canonical formalism, the so-called spin foam models were
proposed as the path integral formalism of LQG \cite{Rovelli}.
However, whether the two approaches are equivalent to each other is
a longstanding open question. Thanks to the development of LQC, we
have much simple models to address this question. As
symmetry-reduced models, there are only finite number of degrees of
freedom in LQC. Following the conventional method in quantum
mechanics, one can find the path integral formalism of LQC starting
with the canonical formalism. This approach is being implemented by
a series of papers \cite{Henderson-1,Huang} with the scheme of
simplified LQC \cite{robust}. In these papers, the transition amplitude were casted
into path integrals by using either \emph{de-parameterized} Hamiltonian or the \emph{group averaging} method.
Here one employed the complete basis
of eigen-states of the volume operator to formulate a path integral
with somehow \emph{descrete-steps}, which inherited certain
properties of spin foams \cite{RV2}. Moreover, the first-order
effective action for the path integral was derived by this approach
\cite{Henderson-1,Huang}, which implied the origin of singularity
resolution of LQC in the path integral representation. In canonical
LQC, the effective Hamiltonian constraint with higher-order quantum
corrections could even be obtained by the semiclassical analysis
using coherent states of Gaussian type, which implied a possible effect of quantum
gravity on large scale cosmology \cite{Ding,Yang}. Moreover, a systematic approach to the effective theories of quantum cosmology in the canonical framework is being developed \cite{BT,bojowald}. It is thus
interesting to see whether the effective Hamiltonian in the canonical theories can be
confirmed by some path integral representation. Since the
higher-order corrections of the Hamiltonian come from the quantum
fluctuations, a natural attempt to achieve them is to employ
coherent state path integral \cite{Brown-book}.

In both WDW quantum cosmology and LQC, by coupling with a massless
scalar field, the Hamiltonian constraint equations can be reformulated
as Klein-Gordon-like equations, where the corresponding gravitational Hamiltonian
operators, as multiplications of several self-adjoint operators, are
non-symmetric in the kinematical Hilbert space \cite{aps1,robust}.
While this treatment is essential in order to obtain the physical
states satisfying the constraint equation, it also provides elegant
physical models to examine the so-called \emph{Moyal}
$*$\emph{-product} in quantum mechanics. At the very beginning,
Moyal proposed the $*$\emph{-product} in order to clarify the role
of statistical concepts in quantum mechanical system \cite{Moyal}.
Then this idea were generalized to many situations including quantum
spacetime itself. It provided the core concept of
\emph{non-commutative geometry}, one of promising and interesting new tools in the study of quantum field theory and
quantum gravity \cite{Connes}. In canonical quantum theories, the $*$\emph{-product} can also be understood by coherent
state approach \cite{book-fuzzy}. In fact, on a coherent state, the expectation value of the multiplication of two
non-commutative self-adjoint operators equals to the $*$\emph{-product} of the expectation values of the two operators.
Thus it is also possible and desirable to derive the $*$\emph{-product} in coherent state
functional integral approach. The above models of quantum cosmology
provide a good arena to launch the desired investigation. Hence we
will study the coherent state functional integrals in spatially flat
isotropic FRW cosmology coupled with a massless scalar field $\phi$
in both WDW and loop
quantization approaches. It should be noted that we will restrict our scheme to Gaussian coherent states for the path integral representation.

\section{Canonical Frameworks of Quantum Cosmology}
The Hilbert-Einstein action for this model is given by:
 \be\label{eqn:action}
 S=\frac{1}{16\pi{G}}\int{d^4x}\sqrt{-g}R
 +\frac{1}{2}\int{d^4x}\sqrt{-g}g^{\mu\nu}\phi_{,\nu}\phi_{,\nu}.
 \ee
In this spatially flat model, we fix a space-like sub-manifold $S$,
which is topologically $\mathbb{R}^3$ and equipped with Cartesian
coordinates $x^i(i=1,2,3)$, and a fiducial flat metric
${}^o\!q_{ab}$ given by ${}^o\!q_{ab}dx^adx^b=dx_1^2+dx_2^2+dx_3^2$.
The physical 3-metric $q_{ab}$ is then determined by
a scale factor $a$ satisfying $q_{ab}=a^2{}^o\!q_{ab}$. It is
convenient to introduce an elementary cell $\mathcal{V}$ and
restrict all integrations to this cell since the spatial slice is non-compact. The
volume of $\mathcal{V}$ with respect to ${}^o\!q_{ab}$ is denoted as
$V_o$ and the physical volume is $V=a^3V_o$. Then the geometrical pair $(a,p_a)$ can be used as canonical variables,
where the conjugate momentum satisfies $p_a\propto{a\dot{a}}$.

In order to study the WDW cosmology and LQC on the same footing, we
employ the new canonical variables $(A_a^i,E_i^a)$ in both theories. Due to the homogeneity and isotropy,
we can fix a set of orthonormal cotriad and triad
$({}^o\!\omega_a^i,{}^o\!e^a_i)$ compatible with ${}^oq_{ab}$ and
adapted to $\mathcal{V}$. Then the cotriad $\omega^i_a$ which are orthonormal with respect to physical metric $q_{ab}$
can be written as $\omega^i_a=\chi{a}{}^o\!\omega_a^i$, where $\chi=1$ if $\omega^i_a$ has the same orientation as the
fiducial ${}^o\!\omega_a^i$ and $\chi=-1$ if the orientation is opposite. The basic canonical variables take the simple
form \cite{mathematical structure}
\be
A_a^i=cV_o^{-(1/3)}{}^o\!\omega_a^i, \ \ \ \ E_i^a=p\sqrt{{}^o\!q}V_o^{-(2/3)}{}^o\!e^a_i.
\ee
The dynamical variables are thus reduced
to $(c,p)$ with the Poisson bracket: $\{c,p\}=8\pi G\gamma/3$, where
$\gamma$ is the Barbero-Immirzi parameter. Following the
\emph{$\bar{\mu}$-schem} of "improved dynamics" \cite{aps1},
the regulator $\bar{\mu}$ used in holonomies is given by
$\bar{\mu}=\sqrt{{\D}/{|p|}}$, where
$\D=4\sqrt{3}\pi\gamma{\ell}_{\textrm{p}}^2$ is a minimum nonzero
eigenvalue of the area operator \cite{Ash-view}. In order to do the
semiclassical analysis, it is convenient to introduce new
dimensionless conjugate variables \cite{robust,Yang}:
 \be\label{eqn:b and v}
 b:=\frac{\bar{\mu}c}{2}=-\frac{4\cdot3^{\frac{1}{4}}G(\pi\gamma)^{\frac{3}{2}}\ell_{\rm{p}}}{3V_o}\frac{p_a}{a^2}
 ,\quad
 v:=\frac{\text{sgn}(p)|p|^{\frac{3}{2}}}{2\pi\gamma{\ell}^2_{\textrm{p}}\sqrt{\D}}
 =\frac{\chi{a^3}V_o}{4\cdot3^{\frac{1}{4}}(\pi\gamma)^{\frac{3}{2}}\ell^3_{\rm{p}}},
 \ee
with the Poisson bracket $\{v,b\}=\-\frac{1}{\hbar}$ , where the
Planck length $\ell_{\textrm{p}}$ is given by
$\ell_{\textrm{p}}^2=G\hbar$. From the matter part of action
(\ref{eqn:action}), we can get the momentum of $\phi$ as
$p_{\phi}=\frac{a^3V_o\dot{\phi}^2}{2}$ and the Poisson bracket:
$\{\phi,p_{\phi}\}=1$. The kinematical Hilbert space of the quantum
theory is supposed to be a tensor product of the gravitational and
matter parts. In WDW quantum cosmology, one employed the standard
\emph{Schr\"{o}dinger} representation for both matter and gravity to
construct Hilbert space $\Hkw$. The (generalized) orthonormal basis
in $\Hkw$ is given by $\ket{v}\otimes\ket{\phi}$ (or denoted as
$\ket{v,\phi}$) with the inner product:
$\inner{v^{\prime},\phi^{\prime}}{v,\phi}=\delta({v^{\prime},v})\delta(\phi^{\prime},\phi)$.
The fundamental operators act on a quantum state
$\Psi(v,\phi)\in\Hkw$ as in standard quantum mechanics. To obtain
the physical states, one has to solve the quantum Hamiltonian
constraint equation \cite{robust}:
 \be\label{eqn:quantum constrain WDW}
 -\CW\cdot\Psi(v,\phi)=
 \left(-\frac{\hat{p}^2_{\phi}}{\hbar^2}+\TW\right)\Psi(v,\phi)
 =\left(\partial^2_{\phi}-12\pi{G}\cdot{v}\partial_v\big(v\partial_v\big)\right)\Psi(v,\phi)=0.
 \ee
However, in LQC, while the \emph{Schr\"{o}dinger} representation was
still used for the matter, gravity was quantized by the polymer-like
representation \cite{mathematical structure}. Thus quantum states in
the gravitational Hilbert space of LQC are functions
$\underline{\Psi}(v)$ of $v$ with support on a countable number of
points and with finite norm
$\norm{\underline{\Psi}}^2:=\sum_{v}|\underline{\Psi}(v)|^2$
\cite{shadow}. Hence the inner product is defined by a
\emph{Kronecker delta} $\inner{v'}{v}=\delta_{v',v}$. The basic
operators act on a quantum state $\underline{\Psi}(v,\phi)$ in the
kinematical Hilbert space $\Hkl$ as:
\be
 \hat{v}\underline{\Psi}(v,\phi)=v\underline{\Psi}(v,\phi),\ \ \ \ \widehat{e^{ib}}\underline{\Psi}(v,\phi)=\underline{\Psi}(v+1,\phi).
 \ee
The quantum Hamiltonian constraint equation becomes
 \be\label{eqn:quantum constrain LQC}
 -\CL\cdot\underline{\Psi}(v,\phi)=\left(-\frac{\hat{p}_{\phi}^2}{\hbar^2}+\TL\right)\underline{\Psi}(v,\phi)=0,
 \ee
where $\TL$ is a positive second-order difference operator defined by a simplified scheme
as \cite{robust}:
 \be\label{eqn:theta action}
 \TL\cdot\underline{\Psi}(v,\phi)=-\frac{3\pi G}{4}
 v\left[(v+2)\underline{\Psi}(v+4,\phi)-2v\underline{\Psi}(v,\phi)+(v-2)\underline{\Psi}(v-4,\phi)\right].
 \ee
Solutions to the constraint equations and their physical inner
products can be obtained through the group averaging procedure. It is demonstrated in \cite{Henderson-1} that in a timeless framework the most important entity is the amplitude
\be
A(v_f,\phi_f;v_i,\phi_i):=\int_{-\infty}^{\infty}d\a\opelem{v_f,\phi_f}{e^{i\a\hat{C}}}{v_i,\phi_i},
\ee
which contains all the dynamical information.

We are going to concern about \emph{coherent state functional integrals}. The
(\emph{generalized}) coherent state of the matter part is labeled by
a complex variable
$z_o:=\frac{1}{\sqrt{2}\c}(\phi_o+\frac{i}{\hbar}\c^2p_{\phi_o})$
and defined by
 \be\label{eqn:coherent state matter}
 \ket{\com{o}}:=\int_{-\infty}^{\infty} d\phi\
 e^{-\frac{(\phi-\phi_o)^2}{2\c^2}}e^{\frac{i}{\hbar}p_{\phi_o}(\phi-\phi_o)}\ket{\phi},
 \ee
which is the eigenstate of the \emph{annihilation} operator
$\hat{z}=\frac{1}{\sqrt{2}\c}(\hat{\phi}+\frac{i}{\hbar}\,\hat{p}_{\phi}\c^2)$,
where $\c$ describes the width of the wave-packet or quantum
fluctuation. It satisfies the key properties of a coherent state,
namely, saturation of Heisenberg's uncertainty relation, resolution
of identity and peakness property. Similarly, we can introduce
$\eta_o:=\frac{1}{\sqrt2\d}(v_o+ib_o\d^2)$ labeling the
coherent state of the gravitational part of WDW theory, which is
defined by
\be
\ket{\cow{o}}:=\int_{-\infty}^{\infty} dv\
 e^{-\frac{(v-v_o)^2}{2\d^2}}e^{-ib_o(v-v_o)}\ket{v}.
 \ee
It also has the analogous properties of the coherent state of matter
part, especially the resolution of the identity:
 \be\label{eqn:WDW identity of coherent}
 \int_{-\infty}^{\infty}dv_o\int_{-\infty}^{\infty}\frac{db_{o}}{2\pi}
 \frac{\ket{\cow{o}}\bra{\cow{o}}}{\inner{\cow{o}}{\cow{o}}}=\mathbb{I}.
 \ee
The whole coherent state of WDW theory reads
$\ket{\com{o}}\ket{\cow{o}}\equiv
\ket{\com{o}}\otimes\ket{\cow{o}}$. On the other hand, due to the
\emph{polymer-like} structure, the coherent state of LQC is
different from that of WDW. Here one can define
$\z_o=\frac{1}{\sqrt2d}(v_o+ib_od^2)$ to label the
\emph{generalized} coherent state \cite{shadow,Ding}:
\be
\obra{\cog{o}}:=\sum_{v\in\mathbb{R}}
 e^{-\frac{(v-v_o)^2}{2d^2}}e^{-ib_o(v-v_o)}\obra{v},
 \ee
where $d$ is the characteristic \emph{width} of the wave packet and
$1\ll{d}\ll{v_o}$ because of the semiclassical feature. For
practical use, one defines the projection of this state on some
lattice of variable $v$, saying the \emph{shadow state}
\cite{shadow}:
 \be\label{eqn:shadow}
 \ket{\cog{o}}^{\rm shad}:=\sum_{k=-\infty}^{\infty}
 e^{-\frac{(k-v_o)^2}{2d^2}}e^{ib_o(k-v_o)}\ket{k},\quad
 k\in\mathbb{Z},
 \ee
where we chose the regular lattice $\{v=k,k\in\mathbb{Z}\}$. This
shadow state also has the analogous properties of a coherent state. Note that our final result will not depend on the particular choice of regular lattice, since the "Hamiltonian operator" in \eqnref{eqn:theta action}
is a difference operator with step of size ``4". The resolution of identity now reads
 \be\label{eqn:gravity identity of coherent}
 \int_{-\infty}^{\infty}dv_o\int_{-\pi}^{\pi}\frac{db_o}{2\pi}
 \frac{\ket{\cog{o}}\bra{\cog{o}}}{\inner{\cog{o}}{\cog{o}}}
 =\sum_{k=-\infty}^{\infty}\ket{k}\bra{k}\equiv\mathbb{I},
 \ee
where the identity $\mathbb{I}$ is in the subspace in which the
states \footnote{Hereafter, without confusion of notations we omit the superscript "shad" of the shadow state for convenience.}
have support only on the regular lattice. It should be noticed that the states with support on semilattice only in a single \emph{semi-axis} of the real line was studied in Ref. \cite{semilattice}, which are superselected by an alternative Hamiltonian constraint operator suitable to deal with some delicate issues in LQC. With this Hamiltonian operator of special symmetrized ordering, the singularity decouples in the kinematical Hilbert space and hence can be removed. It would be interesting to study also the effective equations of this Hamiltonian operator by coherent states supported on semilattice. However, we remark that, for practical calculations in our coherent state functional integral, the dynamical difference equation in \cite{semilattice} still needs to be suitably ``simplified".

\section{Coherent states functional integrals}
In the path integral of the conventional non-relativistic quantum
mechanics, one needs to compute the matrix element of the
\emph{evolution operator} $e^{-i\Delta{t}\hat{H}}$ within the time
interval $\Delta{t}$. However, the situation of cosmology of GR is
very different since both WDW cosmology and LQC are totally
constrained systems, and the operator $\hat{C}$ is not a \emph{true
Hamiltonian}. Instead, we start from the physical inner product,
i.e., the transition amplitude, of coherent states with
\emph{normalization}:
 \be\label{eqn:physical inner coherent}
  A([\Psi_{f}],[\Psi_{i}])\equiv\frac{\bra{\cow{f}}\bra{\com{f}} \int_{-\infty}^{\infty} d\a\ e^{i\a\hat{C}}\ket{\com{i}}\ket{\cow{i}}}
 {\norm{\Psi_{\eta_f}}\norm{\Psi_{z_f}}\norm{\Psi_{z_i}}\norm{\Psi_{\eta_i}}}.
  \ee
To calculate the \emph{transition amplitude}, we split a fictitious
\emph{time interval} $\Delta{\tau}=1$ into $N$ pieces
$\e=\frac{1}{N}$ and thus get
$e^{i\a\hat{C}}=e^{i\sum_{n=1}^{N}\e\a\hat{C}}=\prod_{n=1}^{N}e^{i\e\a\hat{C}}$.
Inserting $N$ times of coherent states resolution of identity of
$\ket{\com{o}}$ and \eqnref{eqn:WDW identity of coherent} (or
\eqnref{eqn:gravity identity of coherent}), \eqnref{eqn:physical
inner coherent} can be casted into
\be
A([\Psi_{f}],[\Psi_{i}])
 =\int_{-\infty}^{\infty}d{\a}\
 A^{\textrm{matt}}_{\a,N}(\Psi_{z_f},\Psi_{z_i})A^{\textrm{grav}}_{\a,N}(\Psi_{\eta_f},\Psi_{\eta_i}),
\ee
where
  \addtocounter{equation}{1}
 \begin{align}
 A^{\rm matt}_{\a,N}=\int_{-\infty}^{\infty}d\phi_{N-1}\dots d\phi_{1}
 \int_{-\infty}^{\infty} \frac{dp_{\phi_{N-1}}}{2\pi\hbar}\dots\frac{dp_{\phi_1}}{2\pi\hbar}
 \prod_{n=1}^{N}
 \frac{\opelem{\com{n}}{e^{i\e\a\frac{\hat{p}_{\phi}^2}{\hbar^2}}}{\com{n-1}}}{\norm{\com{n}}\norm{\com{n-1}}},
 \label{eqn:A matter}\tag{\theequation a}\\
 A^{\rm grav}_{\a,N}=\int_{-\infty}^{\infty}dv_{N-1}\dots dv_{1}
 \int_{-\pi}^{\pi} \frac{db_{N-1}}{2\pi}\dots\frac{db_1}{2\pi}
 \prod_{n=1}^{N}\frac{\opelem{\cow{n}}{e^{-i\e\a\hat{\Theta}}}{\cow{n-1}}}{\norm{\cow{n}}\norm{\cow{n-1}}},
 \label{eqn:A gravity}\tag{\theequation b}
 \end{align}
with $z_N\equiv z_f,z_0\equiv z_i,\eta_N\equiv\eta_f, \textrm{and}
~\eta_0\equiv\eta_i$. Notice that the characteristic widths $\c$ and
$\d$ at different steps are not necessarily the same. So we have to
denote $\c_n$ and $\d_n$ in the semiclassical states $\ket{\com{n}}$
and $\ket{\cow{n}}$ respectively at the ``n-step". Now the main task
is to calculate the matrix element of the exponential operators on
coherent states. The exponential operator $e^{i\e\a\hat{C}}$ can be
expanded as $1+i\e\a\hat{C}+\mathcal{O}(\e^2)$. For the purpose of a
concise writing, we introduce some intermediate-step notations,
 \be\nonumber
   \overline{p}_{\phi_n} \equiv
  \frac{\c^2_np_{\phi_n}+\c^2_{n-1}p_{\phi_{n-1}}}{\c^2_n+\c^2_{n-1}},\quad
  \overline{\c_n^2} \equiv
  \frac{2\c^2_n\c^2_{n-1}}{\c^2_n+\c^2_{n-1}}\quad.
 \ee
Through a detailed calculation shown in the appendix, we get for the matter part
 \ba
 \prod_{n=1}^{N}\frac{\opelem{\com{n}}{e^{i\e\a\frac{\hat{p}_{\phi}^2}{\hbar^2}}}{\com{n-1}}}
{\norm{\com{n}}\norm{\com{n-1}}}
 =\left(\prod_{n=1}^{N}
 \frac{\inner{\com{n}}{\com{n-1}}}{\norm{\com{n}}\norm{\com{n-1}}}\right)
 \exp\Big[\frac{i\e\a}{\hbar^2}
 \sum_{n=1}^{N}\Big(p_{\phi_{n-1}}^2+\frac{\hbar^2}{\c_n^2+\c_{n-1}^2}\Big)\Big],\label{eqn:A-N matter}
 \ea
where the inner product of two adjacent states is
 \be
 \inner{\com{n}}{\com{n-1}}
 =\sqrt\pi\sqrt{\overline{\c_n^2}}
 \exp{\left[-\frac{(\phi_n-\phi_{n-1})^2}{2(\c^2_n+\c^2_{n-1})}
 -\frac{(p_{\phi_n}-p_{\phi_{n-1}})^2\overline{\c_n^2}}{4\hbar^2}
 +\frac{i}{\hbar}\overline{p}_{\phi_n}(\phi_n-\phi_{n-1})\right]},
 \ee
and hence the product of series $\inner{\com{n}}{\com{n-1}}$ can be
expressed as
 \ba
 &&\prod_{n=1}^{N}
 \frac{\inner{\com{n}}{\com{n-1}}}{\norm{\com{n}}\norm{\com{n-1}}}
 =\exp\left[\frac{\phi_N^2+p_{\phi_N}^2\c_{N+1}^2\c_N^2/\hbar^2}{2(\c_{N+1}^2+\c_N^2)}
 -\frac{\phi_0^2+p_{\phi_0}^2\c_{1}^2\c_0^2/\hbar^2}{2(\c_{1}^2+\c_0^2)}\right]
 \left(\prod_{n=1}^{N}\sqrt{\frac{2\c_n\c_{n-1}}{\c_n^2+\c_{n-1}^2}}\right)\nonumber\\
 &&\cdot\exp\Big[\e\sum_{n=1}^{N}\Big(-\frac{2(\c_{n+1}^2+\c_n^2)\phi_n\frac{\phi_n-\phi_{n-1}}{\e}
 -(\c_{n+1}+\c_{n-1})\frac{\c_{n+1}-\c_{n-1}}{\e}\phi_n^2}
 {2(\c_{n+1}^2+\c_{n}^2)(\c_n^2+\c_{n-1}^2)}
 +\frac{i}{\hbar}\overline{p}_{\phi_n}
 \frac{\phi_n-\phi_{n-1}}{\e}\nonumber\\
 &&\quad\quad
 -\frac{1}{4\hbar^2}\frac{4(\c_{n+1}^2\c_n^2\c_{n-1}^2+\c_n^4\c_{n-1}^2)p_{\phi_n}
 \frac{p_{\phi_n}-p_{\phi_{n-1}}}{\e}
 +2\c_{n}^4(\c_{n+1}+\c_{n-1})\frac{\c_{n+1}-\c_{n-1}}{\e}p_{\phi_n}^2}
 {(\c_{n+1}^2+\c_n^2)(\c_n^2+\c_{n-1}^2)}\Big)\Big].\label{eqn:prod
 matter}
 \ea
Here we introduced a \emph{virtual width} $\c_{N+1}$ by hand,
satisfying $\c_{N+1}-\c_{N}=\c_{N}-\c_{N-1}$, in order to get the
tidy sum in the exponential position. In the limit of
$N\rightarrow\infty$, $\c_{N+1}$ will approach $\c_N\equiv\c_f$ and
hence does not effect the quantum dynamics.

For the gravitational part, a careful calculation in WDW quantum cosmology shown in the appendix yields
 \ba
 &&\prod_{n=1}^{N}\frac{\opelem{\cow{n}}{e^{i\e\a\TW}}{\cow{n-1}}}
{\norm{\cow{n}}\norm{\cow{n-1}}}
 =\left(\prod_{n=1}^{N}
 \frac{\inner{\cow{n}}{\cow{n-1}}}{\norm{\cow{n}}\norm{\cow{n-1}}}\right)
 \exp\Big[-i\e\a\cdot12\pi{G}
 \sum_{n=1}^{N}\Big(\big(\frac{1}{\d^2_{n-1}}+b^2_{n-1}\big)
 \big(\overline{v}^2_n+\frac{\overline{\d^2_n}}{2}\big)
 -\frac{\overline{v}^2_n\overline{\d^2_n}}{2\d^4_{n-1}}\nonumber\\
 &&\quad\quad\quad\quad\quad\quad\quad\quad\quad\quad\quad\quad\quad\quad\quad\quad\quad\quad\quad\quad
 \quad\quad\quad\quad\quad\quad
 +\big(\frac{\overline{\d^2_n}}{2\d^2_{n-1}}-\frac{3\big(\overline{\d^2_n}\big)^2}{4\d^4_{n-1}}\big)
 -ib_{n-1}\overline{v}_n\big(\frac{2\overline{\d^2_n}}{\d^2_{n-1}}-1\big)\Big)\Big],\label{eqn:A-N
WDW}
 \ea
where $\overline{v}_n\equiv\frac{\d^2_{n-1}v_n+\d^2_{n}v_{n-1}}{\d^2_n+\d^2_{n-1}},
 \overline{b}_n\equiv\frac{\d^2_nb_n+\d^2_{n-1}b_{n-1}}{\d^2_n+\d^2_{n-1}},
 \overline{\d^2_n}\equiv\frac{2\d^2_n\d^2_{n-1}}{\d^2_n+\d^2_{n-1}}$, and $\prod_{n=1}^{N}\frac{\inner{\cow{n}}{\cow{n-1}}}
{\norm{\cow{n}}\norm{\cow{n-1}}}$ takes the form similar to
\eqnref{eqn:prod matter}.
Now we take the limit $N\rightarrow\infty$
and substitute $\int_{0}^{1}d\tau$ for $\sum_{n=1}^{N}\e$ to get the
functional integral formalism of the amplitude:
 \be\label{eqn:amplitude result WDW}
 A([\Psi_f][\Psi_i])=e^{\frac{1}{2}\left(|z_f|^2-|z_i|^2+|\eta_f|^2-|\eta_i|^2\right)}
 \int{d\a}\int[\mathcal{D}\phi(\tau)][\mathcal{D}p_{\phi}(\tau)][\mathcal{D}v(\tau)][\mathcal{D}b(\tau)]
 e^{i(S_{\a}^{\rm{matt}}+S_{\a}^{\rm{grav}})},
 \ee
where \be\label{eqn:effective action matter}
 S_{\a}^{\rm{matt}}=\int_{0}^{1}d\tau\left(i\frac{d}{d\tau}\left(\frac{\phi^2}{4\c^2}\right)
 +i\frac{d}{d\tau}\left(\frac{\c^2p_{\phi}^2}{4\hbar^2}\right)+\frac{p_{\phi}\dot{\phi}}{\hbar}
 +\frac{\a}{\hbar^2}\left(p_{\phi}^2+\frac{\hbar^2}{2\c^2}\right)\right),
 \ee
\be\label{eqn:effective action WDW}
 S_{\a}^{\rm{grav}}=\int_{0}^{1}d\tau\left(i\frac{d}{d\tau}\left(\frac{v^2}{4\d^2}\right)
 +i\frac{d}{d\tau}\left(\frac{\d^2b^2}{4}\right)-{b\dot{v}}
 -\a12\pi{G}\left[\Big(v^2+\frac{\d^2}{2}\Big)\Big(b^2+\frac{1}{2\d^2}\Big)-ivb\right]\right).
 \ee
Here the "dots" over $\phi$ and $v$ stand for the \emph{time
derivative} with respect to the \emph{fictitious time} $\tau$. The
\emph{functional measures} are defined on \emph{continuous paths} by
taking the limit of $N\rightarrow\infty$:
\begin{subequations}
\ba
 &&\int[\mathcal{D}\phi(\tau)][\mathcal{D}p(\tau)]:=\lim_{N\rightarrow\infty}
 \left(\prod_{n=1}^{N}\sqrt{\frac{2\c_n\c_{n-1}}{\c_n^2+\c_{n-1}^2}}\right)
 \int\prod_{n=1}^{N-1}\frac{d{\phi_n}dp_{\phi_n}}{2\pi\hbar}\ ,\label{eqn:measure-matter}\\
                 &&\int[\mathcal{D}v(\tau)][\mathcal{D}b(\tau)]:=\lim_{N\rightarrow\infty}
 \left(\prod_{n=1}^{N}\sqrt{\frac{2\d_n\d_{n-1}}{\d_n^2+\d_{n-1}^2}}\right)
 \int\prod_{n=1}^{N-1}\frac{d{v_n}db_n}{2\pi}\ .\label{eqn:measure-grav}
 \ea
  \end{subequations}
Ignoring the total derivatives with respect to $\tau$ in Eqs.
(\ref{eqn:effective action matter}) and (\ref{eqn:effective
action WDW}), we can read out the total effective Hamiltonian
constraint in WDW quantum cosmology as:
 \be\label{eqn:effective constraint WDW}
 {\mathscr{H}}_{\rm{eff}}=-\frac{p_{\phi}^2}{\hbar^2}-\frac{1}{2\c^2}
 +12{\pi}G\left[\Big(v^2+\frac{{\d}^2}{2}\Big)
 \Big(b^2+\frac{1}{{2\d}^2}\Big)-ivb\right].
 \ee
Note that $\frac{\d^2}{2}$ and $\frac{1}{2\d^2}$ are the square of
fluctuations of $\hat{v}$ and $\hat{b}$ respectively. They can be
seen as quantum corrections to the leading term: $v^2b^2-ivb$.

On the other hand, careful calculations in LQC shown also in the appendix give
 \ba
 &&\prod_{n=1}^{N}\frac{\opelem{\cog{n}}{e^{i\e\a\TL}}{\cog{n-1}}}
 {\norm{\cog{n}}\norm{\cog{n-1}}}
 =\left(\prod_{n=1}^{N}
 \frac{\inner{\cog{n}}{\cog{n-1}}}{\norm{\cog{n}}\norm{\cog{n-1}}}\right)\nonumber\\
 &\cdot&\exp\Big[-i\e\a{3\pi{G}}\sum_{n=1}^{N}\Big(\Big({\overline{v}_n}^2 +\frac{\overline{d^2_n}}{2}\Big)
 \Big(\sin^2(2\overline{b}_n)\big(1-\frac{8}{d^2_n+d^2_{n-1}}\big)
 +\frac{4}{d_n^2+d_{n-1}^2}\Big)\nonumber\\
 &&\quad\quad\quad\quad\quad\quad\quad\quad\quad\quad\quad\quad\quad\quad\quad\quad\quad\quad\quad\quad\quad\quad\quad
 +i\sin{(4\overline{b}_n)}\overline{v}_n\frac{2d^2_n}{d^2_n+d^2_{n-1}}
 \big(1-\frac{8}{d^2_n+d^2_{n-1}}\big)\Big)\Big],
 \label{eqn:A-N LQC}
 \ea
 where $\overline{v}_n\equiv\frac{d^2_{n-1}v_n+d^2_{n}v_{n-1}}{d^2_n+d^2_{n-1}},
 \overline{b}_n\equiv\frac{d^2_nb_n+d^2_{n-1}b_{n-1}}{d^2_n+d^2_{n-1}},
 \overline{d^2_n}\equiv\frac{2d^2_nd^2_{n-1}}{d^2_n+d^2_{n-1}}$, and $\prod_{n=1}^{N}\frac{\inner{\cog{n}}{\cog{n-1}}}
{\norm{\cog{n}}\norm{\cog{n-1}}}$ also takes the form similar to
\eqnref{eqn:prod matter}.
It should be noted that, in the above calculation, the inner product of two
shadow states is
 \ba
 \inner{\cog{n}}{\cog{n-1}}&=&\sum_{k^{'}}e^{-\frac{(k'-v_n)^2}{2d^2_n}-ib_n(k'-v_n)}\bra{k'}
 \sum_{k}e^{-\frac{(k-v_{n-1})^2}{2d^2_{n-1}}+ib_{n-1}(k-v_{n-1})}\ket{k}\nonumber\\
 &=&\exp{\left[-\frac{(v_n-v_{n-1})^2}{2(d^2_n+d^2_{n-1})}+i\overline{b}_n(v_n-v_{n-1})\right]}
 \sum_k\exp{\left[-\frac{(k-\overline{v}_n)^2}{\overline{d^2_n}}
 -i(b_n-b_{n-1})(k-\overline{v}_n)\right]}\nonumber.
 \ea
Using the so-called \emph{Poisson re-sum formula}:
\be\label{eqn:poisson sum}
 \sum_{k=-\infty}^{\infty}g(k+x)=\sum_{k=-\infty}^{\infty}
 e^{i2\pi{k}x}\int_{-\infty}^{\infty}dy\
 g(y)e^{-i2\pi{k}y},\quad{k\in\mathbb{Z}},
 \ee
we have the summation about $v$ as
 \ba
 &&\sum_k\exp{\left(-(k-\overline{v}_n)^2/\overline{d^2_n}
 -i(b_n-b_{n-1})(k-\overline{v}_n)\right)}\nonumber\\
 &=&\sum_ke^{-i2\pi{k}\overline{v}_n}\int{dy}\
 e^{-y^2/\overline{d^2_n}-i(b_n-b_{n-1})y-i2\pi{k}y}\nonumber\\
 &=&\sqrt\pi\sqrt{\overline{d^2_n}}\sum_ke^{-i2\pi{k}\overline{v}_n}
 \exp{\left(-(b_n-b_{n-1}+2\pi{k})^2\overline{d^2_n}/4\right)}\nonumber
 \ea
and hence the inner product as
 \ba
 \inner{\cog{n}}{\cog{n-1}}\approx\sqrt\pi\sqrt{\overline{d^2_n}}
 \exp{\left(-\frac{(v_n-v_{n-1})^2}{2(d^2_n+d^2_{n-1})}-\frac{(b_n-b_{n-1})^2\overline{d^2_n}}{4}
 +i\overline{b}_n(v_n-v_{n-1})\right)},\label{eqn:inner LQC}
 \ea
where we kept only the $k=0$ term, because in the continuous limit
$N\longrightarrow\infty$, one has $b_n-b_{n-1}\longrightarrow0$ and the
terms corresponding to non-zero integer $k$ are of the same or
higher orders of
$\mathcal{O}\left(e^{-\pi^2\overline{d^2_n}}\right)$ and hence
negligible under the semiclassical condition $d\gg1$. The matrix element $\opelem{\cog{n}}{e^{i\e\a\TL}}{\cog{n-1}}$ can be obtained by similar method.
Thus we obtain:
 \be\label{eqn:amplitude result LQC}
 A([\underline{\Psi}_f][\underline{\Psi}_i])=e^{\frac{1}{2}\left(|z_f|^2-|z_i|^2+|\z_f|^2-|\z_i|^2\right)}
 \int{d\a}\int[\mathcal{D}\phi(\tau)][\mathcal{D}p_{\phi}(\tau)][\mathcal{D}v(\tau)][\mathcal{D}b(\tau)]
 e^{i(\underline{S}_{\a}^{\textrm{matt}}+\underline{S}_{\a}^{\textrm{grav}})}.
 \ee
The effective Hamiltonian constraint in LQC can be read out as:
 \be\label{eqn:effective constraint LQC}
 \underline{{\mathscr{H}}}_{\rm{eff}}=-\frac{p_{\phi}^2}{\hbar^2}-\frac{1}{2\c^2}
 +3{\pi}G\left[\Big(v^2+\frac{d^2}{2}\Big)
 \Big(\sin^2(2b)\big(1-\frac{4}{d^2}\big)+\frac{2}{d^2}\Big)
 +iv\sin{(4b)}\big(1-\frac{4}{d^2}\big)\right],
 \ee
wherein the terms $\frac{d^2}{2}$ and $\frac{2}{d^2}$ are the square
of fluctuations of $\hat{v}$ and $\widehat{\sin(2b)}$ respectively. They are also quantum corrections to the leading term.

At the first sight, both $\mathscr{H}_{\rm{eff}}$ and
$\underline{\mathscr{H}}_{\rm{eff}}$ look problematic due to the
\emph{imaginary part}. One might even suspect the validity of the
coherent state path integral in the models. However, a careful
observation reveals that the real and imaginary parts of the leading
terms can be synthesized into a \emph{Moyal-}$*$ \emph{product}
\cite{book-fuzzy} in both models, i.e.,
\begin{subequations}
\ba
 &&v^2b^2-ivb=ve^{-\frac{i}{2}\left(\overleftarrow{\partial_v}\overrightarrow{\partial}_b
 -\overleftarrow{\partial_b}\overrightarrow{\partial}_v\right)}(bvb)
=:v*(bvb)\label{eqn:WDW*prod},\\
&&v^2\sin^2(2b)+iv\sin{(4b)}=
ve^{\frac{i}{2}\left(\overleftarrow{\partial_v}\overrightarrow{\partial}_b
 -\overleftarrow{\partial_b}\overrightarrow{\partial}_v\right)}\big(\sin{(2b)}v\sin{(2b)}\big)
 =v*\big(\sin{(2b)v}\sin{(2b)}\big)\label{eqn:LQC*prod}.
 \ea
  \end{subequations}
Therefore the effective Hamiltonian constraint in WDW theory takes
the form:
 \be\label{eqn:effective constraint WDW with *}
 {\mathscr{H}}_{\rm{eff}}=-\frac{p_{\phi}^2}{\hbar^2}-\frac{1}{2\c^2}
 +12{\pi}G\left(v*(bvb)+\frac{b^2\d^2}{2}+\frac{v^2}{2\d^2}+\frac{1}{4}\right),
 \ee
while that in LQC becomes
 \be\label{eqn:effective constraint LQC with *}
 \underline{{\mathscr{H}}}_{\rm{eff}}=-\frac{p_{\phi}^2}{\hbar^2}-\frac{1}{2\c^2}
 +3{\pi}G\left(v*\big(\sin{(2b)v}\sin{(2b)}\big)\big(1-\frac{4}{d^2}\big)+\frac{\sin^2{(2b)}d^2}{2}\big(1-\frac{4}{d^2}\big)
 +\frac{2v^2}{d^2}+1\right).
 \ee
To understand how the \emph{Moyal} $*$-\emph{product} emerges in
the gravitational part of the Hamiltonian, recall that both
$\TW\propto\hat{v}(\hat{b}\hat{v}\hat{b})$ and
$\TL\propto\hat{v}(\widehat{\sin{(2b)}}\hat{v}\widehat{\sin{(2b)}})$
are non-symmetric operators which can be regarded as a product of
two self-adjoint operators. Thus, the coherent state functional
integrals suggest the \emph{Moyal} $*$-\emph{product} to express the
effective Hamiltonian for the quantum system with a
\emph{non-symmetric} Hamiltonian operator. Now we explore the motivation of a non-symmetric gravitational Hamiltonian operator through the LQC prescription. It should be noted that the initial Hamiltonian constraint operator in LQC is actually self-adjoint in the kinematical Hilbert space \cite{aps1,robust}. To resolve the constraint equation and find \emph{physical} states, one feasible method is to rebuild the constraint equation as a \emph{Klein-Gordon}-like equation and treat the scalar $\phi$ as an \emph{internal time}. As a result, the constrained quantum system was recast into an unconstrained system of non-relativistic particle whose dynamics is govern by a \emph{Klein-Gordon}-like equation with an \emph{emergent time} variable \cite{aps1,robust}. The price to get this Klein-Gordon-like equation is that the new gravitational Hamiltonian operator $\TL$ becomes a multiplication of two self-adjoint operators, and hence it is no longer symmetric. But this does not indicate that one could not employ $\hat{\Theta}$ in the intermediate step to find physical states. On the other hand, this non-symmetric $\TL$ just provides a suitable arena to examine the \emph{Moyal} $*$-\emph{product} from the path integral perspective. The appearance of \emph{Moyal} $*$-product in our
path integral formalism indicates a possible duality between the path integral formulation on a non-commutative Moyal plane (see e.g. Ref.\cite{path-integral-moyal-plane}) and the canonical quantization on an usual phase space with a "non-symmetric"
Hamiltonian operator.

We can also take another practical way to symmetrize $\TL$ at the beginning. For example, one can define a symmetric version of $\TL$ by
\ba \label{STL}
\TL^{'}:=\frac{1}{2}(\TL+\TL^{\dag})
\propto[\hat{v}(\widehat{\sin{(2b)}}\hat{v}\widehat{\sin{(2b)}})+(\widehat{\sin{(2b)}}\hat{v}\widehat{\sin{(2b)}})\hat{v}],
 \ea
 and then carry out the same procedure of above coherent state functional integral. In the calculation of matrix element $\opelem{\cog{n}}{\TL^{'}}{\cog{n-1}}$, we could think that the operators $\hat{v}$ and $\widehat{\sin{(2b)}}\hat{v}\widehat{\sin{(2b)}}$ in $\TL$ act on \emph{bra} $\bra{\cog{n}}$ and \emph{ket} $\ket{\cog{n-1}}$ respectively, while $\widehat{\sin{(2b)}}\hat{v}\widehat{\sin{(2b)}}$ and $\hat{v}$ in $\TL^{\dag}$ act on \emph{bra} $\bra{\cog{n}}$ and \emph{ket} $\ket{\cog{n-1}}$ respectively. Then it is not difficult to see that the imaginary parts generated by $\TL$ and $\TL^{\dag}$ cancel each other. Hence for the symmetric Hamiltonian operator corresponding to $\TL$, we can get a real effective Hamiltonian constraint without imaginary part for LQC. Similarly, we can also get a real effective Hamiltonian constraint in WDW quantum cosmology by employing a symmetric version of $\TW$.

\section{On the Effective Dynamics}
Using the effective Hamiltonian constraints $\mathscr{H}_{\rm{eff}}$ and $\underline{\mathscr{H}}_{\rm{eff}}$ which contain \emph{Moyal $*$-product}, one may derive the corresponding dynamical equations. For LQC, we can define the evolution equations by:
 \be\label{eqn:evolution with *}
 \dot{f}(v,b):=\frac{1}{i\hbar}\left(f*\underline{\mathscr{H}}_{\rm{eff}}-\underline{\mathscr{H}}_{\rm{eff}}*f\right),
 \ee
for any dynamical quantity $f(v,b)$. Especially, the evolution of basic variables can be obtained as:
 \begin{subequations}
 \ba
 \dot{v}=\frac{12\pi{G}}{\hbar}&\Big[&v*\big(v\sin{(2b)}\cos{(2b)}(1-4\underline{\varepsilon}^2)\big)
 +\frac{\sin{(2b)}\cos{(2b)}(1-4\underline{\varepsilon}^2)}{2\underline{\varepsilon}^2}\nonumber\\
 &&\quad+\left(\frac{v^2}{2}-\Big(v^2+\frac{1}{2\underline{\varepsilon}^2}\Big)\sin^2{(2b)}
 -\frac{\sin^2{(2b)}(1-4\underline{\varepsilon}^2)}{8\underline{\varepsilon}^4}\right)
 \partial_b\underline{\varepsilon}^2\Big],\label{eqn:v evolution with * LQC}\\
 \dot{b}=-\frac{3\pi{G}}{\hbar}&\Big[&\Big(2v\big(1-4\underline{\varepsilon}^2\big)\sin{(2b)}\Big)*\sin{(2b)}
 +{4v}\underline{\varepsilon}^2\nonumber\\
 &&\quad-\frac{1}{\underline{\varepsilon}^4}\Big(\frac{\sin^2{(2b)}}{2}\big(1-{4}\underline{\varepsilon}^2\big)
 +\big(v^2+\frac{1}{2\underline{\varepsilon}^2}\big){4\sin^2{(2b)}}\underline{\varepsilon}^4
 -{2v^2}\underline{\varepsilon}^4\Big)\partial_v\underline{\varepsilon}^2\Big],\label{eqn:b evolution with * LQC}
 \ea
 \end{subequations}
where $\underline{\varepsilon}\equiv1/d$ denote the quantum fluctuation of
$\sin b$, $\partial_b\underline{\varepsilon}^2\equiv \partial (\underline{\varepsilon}^2)/\partial b$ and $\partial_v\equiv \partial/\partial v$. Similarly, we can use the effective
Hamiltonian (\ref{eqn:effective constraint WDW with *}) in WDW
quantum cosmology to get the evolution of basic variables as:
 \begin{subequations}
 \ba
 &&\dot{v}
 =-\frac{12\pi{G}}{\hbar}\left[2v*\big(vb\big)+\frac{b}{\varepsilon^2}
 +\left(\frac{v^2}{2}-\frac{b^2}{2\varepsilon^4}\right)\partial_b\varepsilon^2\right],\label{eqn:v evolution with * WDW}\\
 &&\dot{b}
 =\frac{12\pi{G}}{\hbar}\left[\big(2vb\big)*b+{v}\varepsilon^2
 -\frac{1}{\varepsilon^4}\left(\frac{b^2}{2}-\frac{v^2\varepsilon^4}{2}\right)\partial_v\varepsilon^2\right],
 \label{eqn:b evolution with * WDW}
 \ea
 \end{subequations}
  where ${\varepsilon}\equiv 1/\delta$ denotes the quantum fluctuation of $b$. However, there seem no way to understand Eqs. (\ref{eqn:v evolution with * LQC})-(\ref{eqn:b evolution with * WDW}) directly as effective classical equations because of the $*$-product therein. To get physically predictable effective equations, we have to appeal to other possibilities.

Since the \emph{Moyal $*$-product} originates from the
non-commutativity of operators, one can symmetrize the operator $\TL$ as \eqnref{STL} and repeat the procedure of coherent state functional integrals in last section. Then it is not difficult to get the effective Hamiltonian constraint for LQC as:
 \be\label{eqn:effective constraint LQC remove*}
 \underline{{\mathcal{H}}}:=-\frac{p_{\phi}^2}{\hbar^2}-\frac{1}{2\c^2}
 +3{\pi}G\Big(v^2+\frac{d^2}{2}\Big)\Big(\sin^2{(2b)}\big(1-\frac{4}{d^2}\big)+\frac{2}{d^2}\Big),
 \ee
 which takes the same form as Eq.(\ref{eqn:effective constraint LQC with *}) but without the $*$-product. Note that this effective Hamiltonian constraint is different from that obtained in Ref.\cite{Ding} where a different Hamiltonian constraint operator was employed.
Using the conventional Poisson bracket, we can get the evolution of $v$ as
 \ba
 \dot{v}=-\frac{12\pi{G}}{\hbar}
 \left[\left(v^2+\frac{1}{2\underline{\varepsilon}^2}\right)\sin{(2b)}\cos{(2b)}(1-4\underline{\varepsilon}^2)
 +\left(\frac{v^2}{2}-\left(v^2+\frac{1}{2\underline{\varepsilon}^2}\right)\sin^2{(2b)}
 -\frac{\sin^2{(2b)}(1-4\underline{\varepsilon}^2)}{8\underline{\varepsilon}^4}\right)
 \partial_b\underline{\varepsilon}^2\right].\label{eqn:v evolution without * LQC}
  \ea
 Then a modified Friedmann equation can be derived as
 \ba\label{eqn:Fdm eq LQC}
 &&H^2_{\rm{LQC}}\equiv\left(\frac{\dot{a}}{a}\right)^2=\frac{8\pi{G}\rho_{\rm{c}}}{3}\nonumber\\
  &&\cdot\left[\left(1+\frac{1}{2v^2\underline{\varepsilon}^2}\right)\sin{(2b)}\cos{(2b)}(1-4\underline{\varepsilon}^2)
  +\left(\frac{1}{2}-\left(1+\frac{1}{2v^2\underline{\varepsilon}^2}\right)\sin^2{(2b)}-
  \frac{\sin^2{(2b)}(1-4\underline{\varepsilon}^2)}{8\underline{\varepsilon}^2v^2\underline{\varepsilon}^2}\right)
  \partial_b\underline{\varepsilon}^2\right]^2
 \ea
where $\rho_c\equiv\frac{\sqrt3}{32\pi^2G^2\hbar\gamma^3}$ is a
constant. To annihilate $\sin{(2b)}$ and $\cos{(2b)}$ in \eqnref{eqn:Fdm eq LQC},
 we use the constraint equation (\ref{eqn:effective constraint LQC remove*}) to get
 \be\label{eqn:sin and rho}
 \sin^2{(2b)}=\frac{1}{1-4\underline{\varepsilon}^2}
 \left(\frac{\rho}{\rho_{\rm{c}}}\frac{F}{E}
 -2\underline{\varepsilon}^2\right),
 \ee
  where $\rho=\frac{p^2_{\phi}}{2V^2}$ is the density of matter, $F\equiv1+\frac{\hbar^2}{2\c^2p^2_{\phi}}$, and $E\equiv1+\frac{1}{2v^2\underline{\varepsilon}^2}$.
 However, \eqnref{eqn:Fdm eq LQC} looks problematic since it depends on the volume $v$ of the
chosen fiducial cell. This originates from the fact that we have to
use the coherent states peaked on the phase points $(v, b)$ in the
path integral. In the final picture we have to remove the infrared
regulator by letting the cell occupy full spatial manifold. In this
limit, the irrelevant correction terms proportional to
$1/(v\underline{\varepsilon})^2$ could be neglected, while the relevant terms
proportional to $\underline{\varepsilon}^2$ would be kept, since $\underline{\varepsilon}$
was understood as the fluctuation of $\sin b$ which does not depend
on the fiducial cell. We finally get \be\label{eqn:Fdm eq
LQC2}
 H^2_{\rm{LQC}}=\frac{8\pi{G}\rho_{\rm{c}}}{3}
 \left[\pm\sqrt{\left(\frac{\rho}{\rho_{\rm{c}}}F-2\underline{\varepsilon}^2\right)
 \left(1-4\underline{\varepsilon}^2-\left(\frac{\rho}{\rho_{\rm{c}}}F-2\underline{\varepsilon}^2\right)\right)}
 +\left(\frac{1}{2}-\frac{1}{1-4\underline{\varepsilon}^2}
 \left(\frac{\rho}{\rho_{\rm{c}}}F
 -2\underline{\varepsilon}^2\right)\right)\partial_b\underline{\varepsilon}^2\right]^2,
 \ee
where the positive and negative signs correspond to the expanding and contracting universe respectively. Since $\frac{\hbar^2}{2\c^2}$ is the square of fluctuation of $\hat{p}_{\phi}$, one has $\frac{\hbar^2}{2\c^2p^2_{\phi}}\ll1$. If one ignored the higher order quantum corrections, the modified Friedmann equation (\ref{eqn:Fdm eq LQC2}) would be simplified to the well-known form:
$H^2_{\rm{LQC}}=\frac{8\pi{G}\rho}{3}\left(1-\frac{\rho}{\rho_{\rm{c}}}\right)$. However, \eqnref{eqn:Fdm eq LQC2} implies significant departure
from classical GR, which is manifested in the bounce or
\emph{re-collapse} points determined by $H_{\rm{LQC}}=0$. For a contracting universe, the quantum bounce happens when
 \be\label{eqn:H=0 solution}
 \frac{1}{1-4\underline{\varepsilon}^2}
 \left(\frac{\rho}{\rho_{\rm{c}}}F
 -2\underline{\varepsilon}^2\right)=\frac{1}{2}+\frac{1}{2}
 \sqrt{\frac{(1-4\underline{\varepsilon}^2)^2}{(1-4\underline{\varepsilon}^2)^2
 +(\partial_b\underline{\varepsilon}^2)^2}}.
 \ee
Because $\underline{\varepsilon}\ll1$ and $\partial_b\underline{\varepsilon}^2\ll1$,
 the right hand side of \eqnref{eqn:H=0 solution} could be very close to but no bigger than $1$. Hence the so-called quantum bounce of LQC will occur
when $\rho$ increases to $\rho_{\rm{boun}}\approx\rho_{\rm{c}}$. On
the other hand, for an expanding universe, the positive sign should be chosen in \eqnref{eqn:Fdm eq LQC2}. As a result, the Hubble parameter would always keep non-zero unless $\partial_b\underline{\varepsilon}^2$ approaches $0$ asymptotically. Assuming this is the case, in the asymptotic regime we would get
\be\label{eqn:Fdm eq
LQC3}
 H^2_{\rm{LQC}}=\frac{8\pi{G}\rho_{\rm{c}}}{3}
 \left(\frac{\rho}{\rho_{\rm{c}}}F-2\underline{\varepsilon}^2\right)
 \left(1-4\underline{\varepsilon}^2-\left(\frac{\rho}{\rho_{\rm{c}}}F-2\underline{\varepsilon}^2\right)\right).
 \ee
Therefore, under above assumption,
a re-collapse would occur if $\rho$ decrease to $\rho_{\rm{coll}}\approx
2\underline{\varepsilon}^2\rho_{\rm{c}}$, which coincides with the result in the
canonical theory \cite{Ding,Yang}. As pointed out in Ref.
\cite{Ding}, in this case the inferred re-collapse is almost in all probability
as viewed from the parameter space characterizing the quantum
fluctuation $\underline{\varepsilon}$.

For WDW quantum cosmology, one can also symmetrize the operator $\TW$ and repeat the procedure of coherent state functional integrals. Then it is not difficult to get the effective Hamiltonian constraint:
  \be\label{eqn:effective constraint WDW remove*}
 {{\mathcal{H}}}:=-\frac{p_{\phi}^2}{\hbar^2}-\frac{1}{2\c^2}
 +12{\pi}G\Big(v^2+\frac{\d^2}{2}\Big)\Big(b^2+\frac{1}{2\d^2}\Big),
 \ee
 which takes the same form as Eq.(\ref{eqn:effective constraint WDW with *}) but without the $*$-product. Using this Hamiltonian constraint, we can get the modified Friedmann equation for WDW cosmology as:
 \be\label{eqn:Fdm eq WDW}
 H^2_{\rm{WDW}}=\frac{8\pi{G}\rho_{\rm{c}}}{3}
 \left[\pm\sqrt{\frac{\rho}{\rho_{\rm{c}}}F-2{\varepsilon}^2}
 +\frac{\partial_b{\varepsilon}^2}{2}\right]^2,
 \ee
  where the positive and negative signs correspond to the expanding and contracting universe respectively. It is obvious from \eqnref{eqn:Fdm eq WDW} that there would be no bounce for a contracting universe. For an expanding universe, the Hubble parameter $H^2_{\rm{WDW}}$ might vanish only if $\partial_b\varepsilon^2$ approaches $0$ asymptotically. In this case, \eqnref{eqn:Fdm eq WDW} implies that a re-collapse would also happen if the density of matter could decrease to $\rho\approx \rho_{\rm{coll}}=2{\varepsilon}^2\rho_c$.  Hence, once higher-order quantum corrections are included, the
inferred re-collapse is a common effect in both WDW cosmology and
LQC under the condition that quantum fluctuations approach constant asymptotically.
Intuitively, as the universe expands unboundedly, the matter density
would become so tiny that its effect could be comparable to that of
quantum fluctuations of the spacetime geometry. Then the Hamiltonian
constraint may force the universe to contract back.

\section{Concluding Remarks}
The minisuperspace models of quantum cosmology provide the good avena for testing the ideas and constructions of quantum gravity theories. A few
physically significant results have been obtained in both WDW cosmology and LQC. In LQC,
the big bang singularity is resolved by the quantum bounce, and the effective Hamiltonian constraint with higher-order quantum
corrections could even be obtained by the semiclassical analysis, which implied a possible effect of quantum
gravity on large scale cosmology. It is desirable to study such kind of predictions from different perspectives and in different frameworks.  Since the
higher-order corrections of the Hamiltonian come from the quantum
fluctuations, a natural attempt to achieve them is to employ
coherent state path-integral. On the other hand, the so-called \emph{Moyal}
$*$\emph{-product} in quantum mechanics is generalized to many situations including quantum
spacetime itself. It is also possible and desirable to derive the $*$\emph{-product} by coherent state
functional integral approach within quantum cosmological models. These issues have been addressed in previous sections.

We summarize our main results with a few remarks. First, by the well-established canonical theories,
the coherent state functional integrals for both WDW cosmology and LQC have been formulated by group averaging. As far as we know, this is the first attempt to apply coherent state functional integral to the models of quantum cosmology. Second, the main calculation results of our coherent state functional integrals are the effective Hamiltonian constraints (\ref{eqn:effective constraint WDW with *}) and (\ref{eqn:effective constraint LQC with *}) for WDW cosmology and LQC respectively. These show that the \emph{Moyal (star)-product} can emerge naturally in the path integral approach via the
effective Hamiltonian with higher-order quantum corrections. That is the main reason why we start with the non-symmetric gravitational Hamiltonian constraint operators $\TW$ and $\TL$ for the path integrals. Whether the resulted Hamiltonian constraints with $*$-product could make some physical prediction would be an interesting open issue. Tentatively, the appearance of \emph{Moyal} $*$-product in our coherent state
path integral indicates a possible duality between the path integral formulation on a non-commutative Moyal plane and the canonical quantization on an usual phase space with a "non-symmetric"
Hamiltonian operator. Third, for symmetric Hamiltonian constraint operators, the effective theories and modified Friedmann equations have been obtained by the coherent state path integrals in both WDW cosmology and LQC. For LQC, the effective equation (\ref{eqn:Fdm eq
LQC2}) can reduce to the first-order modified Friedmann equation when higher order quantum corrections are neglected. Hence the quantum bounce resolution of big bang singularity can also be obtained by the path integral representation. On the other hand, if higher order corrections are included, under the condition that quantum fluctuations approach constant asymptotically, there is great possibility for the re-collapse of an expanding universe due to the quantum gravity effect, which coincides with the result obtained in canonical LQC. Moreover, the effective equations imply that the inferred effect of re-collapse is common in both
WDW cosmology and LQC under above condition. Finally, it should be noted that,
as we used the coherent states of Gaussian type, the effective equations and hence the inferred effect of re-collapse are only valid with the assumption that these coherent states can faithfully represent the semiclassical behaviors of WDW cosmology and LQC. Whether there is a similar result for other semiclassical states is still an interesting open issue.

\section*{Acknowledgments}
We thank Abhay Ashtekar and Peng Xu for helpful suggestion and
discussion. This work is supported by NSFC (No.10975017) and the
Fundamental Research Funds for the Central Universities.

\appendix

\section{Calculation of the functional integrals}
We will use the following \emph{Gaussian integrals}:
\begin{subequations}
\ba
 \label{eqn:integral original}&&\int_{-\infty}^{+\infty}dx\ e^{-a^2x^2}=\frac{\sqrt\pi}{a},\\
 \label{eqn:integral x2}&&\int_{-\infty}^{+\infty}dx\ x^2e^{-a^2x^2}=\frac{\sqrt\pi}{a}\frac{1}{2a^2},\\
 \label{eqn:integral x4}&&\int_{-\infty}^{+\infty}dx\ x^4e^{-a^2x^2}=\frac{\sqrt\pi}{a}\frac{3}{4a^4},\\
 \label{eqn:integral cos}&&\int_{-\infty}^{+\infty}dx\ \cos{(bx)}e^{-a^2x^2}=
 \frac{\sqrt\pi}{a}\ e^{-\frac{b^2}{4a^2}},\\
 \label{eqn:integral x2cos}&&\int_{-\infty}^{+\infty}dx\ x^2\cos{(bx)}e^{-a^2x^2}=
 \frac{\sqrt\pi}{a}\left(\frac{1}{2a^2}-\frac{b^2}{4a^4}\right)e^{-\frac{b^2}{4a^2}},\\
 \label{eqn:integral xsin}&&\int_{-\infty}^{+\infty}dx\ x\sin{(bx)}e^{-a^2x^2}=
 \frac{\sqrt\pi}{a}\frac{b}{2a^2}\ e^{-\frac{b^2}{4a^2}},\\
 \label{eqn:integral x3sin}&&\int_{-\infty}^{+\infty}dx\ x^3\sin{(bx)}e^{-a^2x^2}=
 \frac{\sqrt\pi}{a}\left(\frac{3b}{4a^4}-\frac{b^3}{8a^6}\right) \ e^{-\frac{b^2}{4a^2}},\\
 \label{eqn:integral x4cos}&&\int_{-\infty}^{+\infty}dx\ x^4\cos{(bx)}e^{-a^2x^2}=
 \frac{\sqrt\pi}{a}\left(\frac{3}{4a^4}-\frac{3b^2}{4a^6}+\frac{b^4}{16a^8}\right) \ e^{-\frac{b^2}{4a^2}},
\ea
\end{subequations}
with constant parameters $a>0$, $b\in\mathbb{R}$. For the simplicity of notation, we denote the momentum conjugate to
$\phi$ by $p$ rather than $p_{\phi}$ in this appendix.

The main task is to calculate the matrix elements of exponentiated
operators:
$\opelem{\com{n}}{e^{i\e\a\frac{\hat{p}^2}{\hbar^2}}}{\com{n-1}}$,
$\opelem{\cow{n}}{e^{-i\e\a{\TW}}}{\cow{n-1}}$ and
$\opelem{\cog{n}}{e^{-i\e\a\TL}}{\cog{n-1}}$ . For matter part, we get
 \ba
 \inner{\com{n}}{\com{n-1}}
 &=&\exp{\left(-\frac{(\phi_n-\phi_{n-1})^2}{2(\c_n^2+\c_{n-1}^2)}
 +\frac{i}{\hbar}\frac{\c^2_np_n+\c^2_{n-1}p_{n-1}}{\c^2_n+\c^2_{n-1}}
 \big(\phi_n-\phi_{n-1}\big)\right)}\nonumber\\
 &&\cdot\int{d\phi}
 \exp{\left[-\frac{\c_n^2+\c_{n-1}^2}{2\c_n^2\c_{n-1}^2}
 \left(\phi-\frac{\c_{n-1}^2\phi_n+\c_n^2\phi_{n-1}}{\c^2_n+\c_{n-1}^2}\right)^2-\frac{i}{\hbar}\big(p_n-p_{n-1}\big)
 \left(\phi-\frac{\c_{n-1}^2\phi_n+\c_n^2\phi_{n-1}}{\c^2_n+\c_{n-1}^2}\right)\right]}.\nonumber
 \ea
Introducing some intermediate-step notations:
 \be\nonumber
  \overline{\phi}_n \equiv
  \frac{\c_{n-1}^2\phi_n+\c_n^2\phi_{n-1}}{\c^2_n+\c_{n-1}^2},\quad
  \overline{p}_n \equiv
  \frac{\c^2_np_n+\c^2_{n-1}p_{n-1}}{\c^2_n+\c^2_{n-1}},\quad
  \overline{\c_n^2} \equiv
  \frac{2\c^2_n\c^2_{n-1}}{\c^2_n+\c^2_{n-1}},
    \ee
we can get a concise writing:
 \ba
 \inner{\com{n}}{\com{n-1}}
 =\exp{\left(-\frac{(\phi_n-\phi_{n-1})^2}{2(\c_n^2+\c_{n-1}^2)}
 +\frac{i}{\hbar}\overline{p}_n(\phi_n-\phi_{n-1})\right)}
 \int{d\phi}\exp{\left(-\frac{(\phi-\overline{\phi}_n)^2}{\overline{\c_n^2}}\right)}
 \exp{\left(-\frac{i}{\hbar}(p_n-p_{n-1})(\phi-\overline{\phi}_n)\right)},\nonumber
 \ea
and the integration wherein is easy to be done by changing the
integrated variable $\phi$ to
$\tilde{\phi}\equiv\phi-\overline{\phi}_n$:
 \be
 \int{d\tilde{\phi}}\cos{\left(-\frac{i}{\hbar}(p_n-p_{n-1})\tilde{\phi}\right)}
 \exp{\left(-\frac{{\tilde{\phi}}^2}{\overline{\c_n^2}}\right)}=\sqrt\pi\sqrt{\overline{\c_n^2}}
 \exp{\left(-\frac{(p_n-p_{n-1})^2\overline{\c_n^2}}{4\hbar^2}\right)}\nonumber.
 \ee
Finally we get the inner product:
 \be
 \inner{\com{n}}{\com{n-1}}
 =\sqrt\pi\sqrt{\overline{\c_n^2}}
 \exp{\left[-\frac{(\phi_n-\phi_{n-1})^2}{2(\c^2_n+\c^2_{n-1})}
 -\frac{(p_n-p_{n-1})^2\overline{\c_n^2}}{4\hbar^2}
 +\frac{i}{\hbar}\overline{p}_n(\phi_n-\phi_{n-1})\right]}.
 \ee
To get \eqnref{eqn:prod matter}, we need to deal with the product
 \ba
 \prod_{n=1}^{N}\frac{\inner{\com{n}}{\com{n-1}}}{\norm{\com{n}}\norm{\com{n-1}}}
  &=&\left(\prod_{n=1}^{N}\sqrt{\frac{2\c_n\c_{n-1}}{\c^2_n+\c^2_{n-1}}}\right)
 \exp{\left[\sum_{n=1}^{N}\frac{i}{\hbar}\overline{p}_n(\phi_n-\phi_{n-1})\right]}\nonumber\\
 &&\quad\quad\cdot\exp{\left[\sum_{n=1}^{N}\left(-\frac{\phi^2_n-2\phi_n\phi_{n-1}}{2(\c^2_n+\c^2_{n-1})}
 -\frac{\phi^2_{n-1}}{2(\c^2_n+\c^2_{n-1})}
 -\frac{(p^2_n-2p_np_{n-1})\overline{\c^2_n}}{4\hbar^2}-\frac{p^2_{n-1}\overline{\c^2_n}}{4\hbar^2}\right)\right]}
 \label{eqn:prod1}
 \ea
where the summation in the last part of exponential in
\eqnref{eqn:prod1} can be re-organized as
 \ba
 &&\sum_{n=1}^{N}
 \left(-\frac{\phi^2_n-2\phi_n\phi_{n-1}}{2(\c^2_n+\c^2_{n-1})}-\frac{\phi^2_{n-1}}{2(\c^2_n+\c^2_{n-1})}\right)
 \nonumber\\
 &=&\frac{\phi^2_N}{2(\c^2_{N+1}+\c^2_N)}
 -\left(\frac{\phi^2_N}{2(\c^2_{N+1}+\c^2_N)}+\frac{\phi^2_N-2\phi_N\phi_{N-1}}{2(\c^2_N+\c^2_{N-1})}\right)-\cdots
 -\left(\frac{\phi^2_1}{2(\c^2_{2}+\c^2_1)}+\frac{\phi^2_1-2\phi_1\phi_{0}}{2(\c^2_1+\c^2_{0})}\right)
 -\frac{\phi^2_{0}}{2(\c^2_1+\c^2_{0})}\nonumber\\
  &=&\frac{\phi^2_N}{2(\c^2_{N+1}+\c^2_N)}-\frac{\phi^2_{0}}{2(\c^2_1+\c^2_{0})}
 -\sum_{n=1}^{N}\frac{2(\c^2_{n+1}+\c^2_{n})\phi_n(\phi_n-\phi_{n-1})
 -(\c_{n+1}+\c_{n-1})(\c_{n+1}-\c_{n-1})\phi^2_n}
 {2(\c^2_{n+1}+\c^2_n)(\c^2_n+\c^2_{n-1})}\label{eqn:sum phi}
 \ea
and similarly
 \ba
 &&\sum_{n=1}^{N}\left(-\frac{(p^2_n-2p_np_{n-1})\overline{\c^2_n}}{4\hbar^2}
 -\frac{p^2_{n-1}\overline{\c^2_n}}{4\hbar^2}\right)
 =\frac{p^2_N\c^2_{N+1}\c^2_{N}}{2\hbar^2(\c^2_{N+1}+\c^2_N)}
 -\frac{p^2_{0}\c^2_1\c^2_0}{2\hbar^2(\c^2_1+\c^2_{0})}\nonumber\\
 &&\quad\quad\quad\quad-\sum_{n=1}^{N}\frac{4(\c^2_{n+1}\c^2_{n}\c^2_{n-1}+\c^4_n\c^2_{n-1})p_n(p_n-p_{n-1})
 +2\c^4_n(\c_{n+1}+\c_{n-1})(\c_{n+1}-\c_{n-1})p^2_n}{4\hbar^2(\c^2_{n+1}+\c^2_n)(\c^2_n+\c^2_{n-1})}.
 \label{eqn:sum p}
 \ea
Collecting the above results of \eqnref{eqn:prod1}, \eqnref{eqn:sum
phi} and \eqnref{eqn:sum p}, we can finally get \eqnref{eqn:prod  matter}. The
matrix element
$\opelem{\com{n}}{{i\e\a\frac{\hat{p}^2}{\hbar^2}}}{\com{n-1}}$ is
proportional to
  \ba
 &&\opelem{\com{n}}{\hat{p}^2}{\com{n-1}}=\exp{\left(-\frac{(\phi_n-\phi_{n-1})^2}{2(\c^2_n+\c^2_{n-1})}
 +\frac{i}{\hbar}\overline{p}_n(\phi_n-\phi_{n-1})\right)}
 \nonumber\\
 &&\cdot\int{d\phi}\left[p_{n-1}^2+\frac{\hbar^2}{\c_{n-1}^2}-\frac{\hbar^2}{\c^4_{n-1}}(\phi-\phi_{n-1})^2
 +\frac{2i\hbar{p_{n-1}}}{\c_{n-1}^2}(\phi-\phi_{n-1})\right]
 \exp{\left(-\frac{(\phi-\overline{\phi}_{n})^2}{\overline{\c^2_n}}
 -\frac{i}{\hbar}(p_n-p_{n-1})(\phi-\overline{\phi}_n)\right)}.\nonumber
 \ea
To do the above integral, we have to rewrite the integrand as a
function of $\phi-\overline{\phi}_n$. Except for the exponential
function, the left terms are polynomials of $\phi$. The integral of
zeroth power term $p^2_{n-1}+\frac{\hbar^2}{\c^2_n}$ gives
 \be\nonumber
 \sqrt\pi\sqrt{\overline{\c^2_n}}
 \exp{\left(-\frac{(p_n-p_{n-1})^2\overline{\c^2_n}}{4\hbar^2}\right)}
 \cdot\left(p^2_{n-1}+\frac{\hbar^2}{\c^2_n}\right),
 \ee
and hence its contribution to matrix element
$\opelem{\com{n}}{\hat{p}^2}{\com{n-1}}$ is:
$\inner{\com{n}}{\com{n-1}}\left(p^2_{n-1}+\frac{\hbar^2}{\c^2_n}\right)$.
To deal with the linear and square terms of $\phi$, we first rewrite
 \ba
 &&\phi-\phi_{n-1}=\phi-\overline{\phi}_n+\overline{\phi}_n-\phi_{n-1}
 =\phi-\overline{\phi}_n+\frac{\c^2_{n-1}}{\c^2_n+\c^2_{n-1}}(\phi_n-\phi_{n-1})
 \equiv\phi-\overline{\phi}_n+\phi^{'}_n,\nonumber\\
&&(\phi-\phi_{n-1})^2=(\phi-\overline{\phi}_n+\overline{\phi}_n-\phi_{n-1})^2
=(\phi-\overline{\phi}_n)^2+2\phi^{'}_n(\phi-\overline{\phi}_n)+{\phi_n^{'}}^2,\nonumber
 \ea
and get the integrals:
 \ba
 &&\int{d\phi}(\phi-\overline{\phi}_n)^2
 \exp{\left(-\frac{(\phi-\overline{\phi}_n)^2}{\overline{\c^2_n}}
 -\frac{i}{\hbar}(p_n-p_{n-1})(\phi-\overline{\phi}_n)\right)}\nonumber\\
  &=&\sqrt\pi\sqrt{\overline{\c^2_n}}
 \exp{\left(-\frac{(p_n-p_{n-1})^2\overline{\c^2_n}}{4\hbar^2}\right)}
 \left(\frac{\overline{\c_n^2}}{2}-\frac{(p_n-p_{n-1})^2(\overline{\c_n^2})^2}{4\hbar^2}\right),\nonumber\\
&&\int{d\phi}(\phi-\overline{\phi}_n)
 \exp{\left(-\frac{(\phi-\overline{\phi}_n)^2}{\overline{\c^2_n}}
 -\frac{i}{\hbar}(p_n-p_{n-1})(\phi-\overline{\phi}_n)\right)}\nonumber\\
  &=&\sqrt\pi\sqrt{\overline{\c^2_n}}
 \exp{\left(-\frac{(p_n-p_{n-1})^2\overline{\c^2_n}}{4\hbar^2}\right)}\cdot
 \frac{-i(p_n-p_{n-1})\overline{\c^2_n}}{2\hbar}\nonumber.
 \ea
Combining all the above results, the matrix element
$\opelem{\com{n}}{e^{i\e\a\frac{\hat{p}^2}{\hbar^2}}}{\com{n-1}}$ is
 \ba
 &&\inner{\com{n}}{\com{n-1}}+\frac{i\e\a}{\hbar^2}\inner{\com{n}}{\com{n-1}}
 \Big(p^2_{n-1}+\frac{\hbar^2}{\c^2_{n-1}}
 -\frac{\hbar^2\overline{\c^2_n}}{2\c^4_{n-1}}
 -\frac{(p_n-p_{n-1})^2(\overline{\c_n^2})^2}{4\c^4_{n-1}}
 -\frac{\hbar^2(\phi^{'}_n)^2}{\c^4_{n-1}}\nonumber\\
 &&\quad\quad\quad\quad\quad\quad\quad\quad\quad\quad\quad\quad\quad\quad\quad
 +\frac{i\hbar\phi^{'}_n\overline{\c^2_n}(p_n-p_{n-1})}{\c^4_{n-1}}
 +\frac{p_{n-1}\overline{\c^2_n}(p_n-p_{n-1})}{\c^2_{n-1}}
 +\frac{2i\hbar{p_{n-1}\phi^{'}_n}}{\c^2_{n-1}}\Big)+\mathcal{O}(\e^2)\nonumber\\
 &&=\inner{\com{n}}{\com{n-1}}
 \exp{\left[\frac{i\e\a}{\hbar^2}\left(p^2_{n-1}
 +\frac{\hbar^2}{\c^2_n+\c^2_{n-1}}+P^{\textrm{matt}}_{n,n-1}\right)\right]},
 \ea
up to $\mathcal{O}(\e^2)$. Here $P^{\rm matt}_{n,n-1}$
denotes a polynomial of $p_n-p_{n-1}$ and $\phi_n-\phi_{n-1}$
without zeroth order terms. To get a functional integral formalism,
one has to take the number of steps $N\longrightarrow\infty$ and so
$\phi_n-\phi_{n-1}\longrightarrow0$, $p_n-p_{n-1}\longrightarrow0$.
Furthermore, there is already an infinitesimal $\e$ multiplied to
$P^{\rm matt}_n$. This means that $P^{\rm matt}_{n,n-1}$ does not
become a \emph{time derivative} in the limit of
$N\longrightarrow\infty$. As a result, $P^{\rm matt}_{n,n-1}$ does
not contribute to \emph{effective action} ${S}_{\a}^{\textrm{matt}}$.

For gravitational part of WDW theory, we get
 \ba
 \opelem{\cow{n}}{e^{-i\e\a{\TW}}}{\cow{n-1}}
 =\inner{\cow{n}}{\cow{n-1}}+i\e\a(12\pi{G})
 \int{dv}\cow{n}^*(v)v\partial_{v}\big[v\partial_v\cow{n-1}(v)\big]+\mathcal{O}(\e^2).\label{eqn:WDW-matrix-element}
 \ea
Similar to the matter part, it is easy to get the inner product of
two coherent states as
 \be
 \inner{\cow{n}}{\cow{n-1}}=\sqrt\pi\sqrt{\overline{\d^2_n}}
 \exp{\left(-\frac{(v_n-v_{n-1})^2}{2(\d^2_n+\d^2_{n-1})}-\frac{(b_n-b_{n-1})^2\overline{\d^2_n}}{4}
 -i\overline{b}_n(v_n-v_{n-1})\right)},\label{eqn:inner WDW}
 \ee
where some notations are defined as before:
 \be
 \overline{b}_n\equiv\frac{\d^2_nb_n+\d^2_{n-1}b_{n-1}}{\d^2_n+\d^2_{n-1}},\quad
 \overline{\d^2_n}\equiv\frac{2\d^2_n\d^2_{n-1}}{\d^2_n+\d^2_{n-1}}.\nonumber
 \ee
 The second term of \eqnref{eqn:WDW-matrix-element} is proportional to
 \ba
 \int{dv}\cow{n}^*(v)v\partial_{v}\big[v\partial_v\cow{n-1}(v)\big]
 =e^{-\frac{(v_n-v_{n-1})^2}{2(\d^2_n+\d^2_{n-1})}-i\overline{b}_n(v_n-v_{n-1})}
 \int{dv}\ e^{-\frac{(v-\overline{v}_n)^2}{\overline{\d^2_n}}+i(b_n-b_{n-1})(v_n-\overline{v}_n)}&&\nonumber\\
 \cdot\left[-\left(\frac{1}{\d^2_{n-1}}+b^2_{n-1}\right)v^2
 +\frac{v^2(v-v_{n-1})^2}{\d^4_{n-1}}+\frac{2ib_{n-1}v^2(v-v_{n-1})}{\d^2_{n-1}}
 -\frac{v(v-v_{n-1})}{\d^2_{n-1}}-ib_{n-1}v\right],&&\label{eqn:WDW-integral}
 \ea
where
$\overline{v}_n\equiv\frac{\d^2_{n-1}v_n+\d^2_nv_{n-1}}{\d^2_n+\d^2_{n-1}}$.
To do the integral, we first have to rewrite the integrand as a
function of $v-\overline{v}_{n}$ as follows:
 \ba
&&-\left(\frac{1}{\d^2_{n-1}}+b^2_{n-1}\right)v^2
 +\frac{v^2(v-v_{n-1})^2}{\d^4_{n-1}}+\frac{2ib_{n-1}v^2(v-v_{n-1})}{\d^2_{n-1}}
 -\frac{v(v-v_{n-1})}{\d^2_{n-1}}-ib_{n-1}v\nonumber\\
 &=&-\left[\left(\frac{1}{\d^2_{n-1}}+b^2_{n-1}\right)\overline{v}^2_n+ib_{n-1}\overline{v}_n
 \right]+\frac{(v-\overline{v}_n)^4}{\d^4_{n-1}}
 +\left[\frac{2(\overline{v}_n+v'_n)}{\d^4_{n-1}}+\frac{2ib_{n-1}}{\d^2_{n-1}}\right]
 (v-\overline{v}_n)^3\nonumber\\
 &&-\left[\left(\frac{1}{\d^2_{n-1}}+b^2_{n-1}\right)
 -\frac{\overline{v}^2_n+(v'_n)^2+4\overline{v}_nv'_n}{\d^4_{n-1}}
 -\frac{2ib_{n-1}(2\overline{v}_n+v'_{n})}{\d^2_{n-1}}+\frac{1}{\d^2_{n-1}}\right](v-\overline{v}_n)^2\nonumber\\
 &&-\left[2\overline{v}_n\left(\frac{1}{\d^2_{n-1}}+b^2_{n-1}\right)
 -\frac{2\overline{v}_nv'_{n}(\overline{v}_n+v'_n)}{\d^4_{n-1}}
 -\frac{2ib_{n-1}(\overline{v}^2_n+2\overline{v}_nv'_n)}{\d^2_{n-1}}
 -\frac{\overline{v}_n+v'_n}{\d^2_{n-1}}+ib_{n-1}\right](v-\overline{v}_n)\nonumber\\
 &&\quad+\left(\frac{\overline{v}^2_n{v'_n}}{\d^4_{n-1}}+\frac{2ib_{n-1}\overline{v}^2_n}{\d^2_{n-1}}
 -\frac{\overline{v}_n}{\d^2_{n-1}}\right)v'_n,
 \ea
where
$v'_n\equiv\overline{v}_n-v_{n-1}=\frac{\d^2_{n-1}(v_n-v_{n-1})}{\d^2_n+\d^2_{n-1}}$.
Doing the integral term by term, we can obtain the result of
\eqnref{eqn:WDW-integral}:
 \ba
 \int{dv}\cow{n}^*(v)v\partial_{v}\big[v\partial_v\cow{n-1}(v)\big]
 &=&-\inner{\cow{n}}{\cow{n-1}}
 \Big[\Big(\frac{1}{\d^2_{n-1}}+b^2_{n-1}\Big)\overline{v}^2_n+ib_{n-1}\overline{v}_n
 -\frac{3\big(\overline{\d^2_n}\big)^2}{4\d^4_{n-1}}\nonumber\\
 &&\quad\quad+\Big(\big(\frac{1}{\d^2_{n-1}}+b^2_{n-1}\big)-\frac{\overline{v}^2_n}{\d^4_{n-1}}
 -\frac{4ib_{n-1}\overline{v}_n}{\d^2_{n-1}}+\frac{1}{\d^2_{n-1}}\Big)\frac{\overline{\d^2_n}}{2}
 +P^{\textrm{WDW}}_{n,n-1}\Big],
 \ea
and finally the matrix element:
 \ba
 \opelem{\cow{n}}{e^{-i\e\a{\TW}}}{\cow{n-1}}&=&\inner{\cow{n}}{\cow{n-1}}\exp\Big[-i\e\a\cdot12\pi{G}
\Big(\big(\frac{1}{\d^2_{n-1}}+b^2_{n-1}\big)\big(\overline{v}^2_n+\frac{\overline{\d^2_n}}{2}\big)
-\frac{\overline{v}^2_n\overline{\d^2_n}}{2\d^4_{n-1}}
 +\Big(\frac{\overline{\d^2_n}}{2\d^2_{n-1}}-\frac{3\big(\overline{\d^2_n}\big)^2}{4\d^4_{n-1}}\Big)\nonumber\\
 &&\quad\quad\quad\quad\quad\quad\quad\quad\quad\quad\quad\quad\quad\quad\quad
 -ib_{n-1}\overline{v}_n\Big(\frac{2\overline{\d^2_n}}{\d^2_{n-1}}-1\Big)+P^{\textrm{WDW}}_{n,n-1}\Big)\Big],
 \ea
where $P^{\textrm{WDW}}_{n,n-1}$ is a polynomial of $v_n-v_{n-1}$,
$b_n-b_{n-1}$ and $\d_n-\d_{n-1}$ without the zeroth order term. As
before, this quantity does not contribute to the effective action
under the continuous limit $\e\equiv1/{N}\longrightarrow0$.

For gravitational part of LQC, we have to calculate the inner
product $\inner{\cog{n}}{\cog{n-1}}$ and matrix element
$\opelem{\cog{n}}{-i\e\a{\TL}}{\cog{n-1}}$. The inner product of two
shadow states has been obtained in \eqnref{eqn:inner LQC}.
The order of $\mathcal{O}(\e)$
of the matrix element $\opelem{\cog{n}}{e^{-i\e\a\TL}}{\cog{n-1}}$
is
 \ba
 \opelem{\cog{n}}{-i\e\a\TL}{\cog{n-1}}&=&i\e\a\frac{3\pi{G}}{4}
 \sum_{k}\Big[k(k+2)\cog{n}^*(k)\cog{n-1}(k+4)
 -2k^2\cog{n}^*(k)\cog{n-1}(k)\nonumber\\
 &&\quad\quad\quad\quad\quad\quad\quad+k(k-2)\cog{n}^*(k)\cog{n-1}(k-4)\Big]\nonumber\\
 &\equiv&i\e\a\frac{3\pi{G}}{4}
 \Big(D^+_{n,n-1}-D^0_{n,n-1}+D^-_{n,n-1}\Big).\label{eqn:app-B-matirx-element}
 \ea
Now we need to deal with the three terms
$D^+_{n,n-1},D^0_{n,n-1},D^-_{n,n-1}$ separately.
First, we get
 \ba
 &&D^+_{n,n-1}\equiv\sum_{k}(k^2+2k)e^{-\frac{(k-v_n)^2}{2d_n^2}-\frac{(k+4-v_{n-1})^2}{2d^2_{n-1}}}
 e^{-ib_n(k-v_n)+ib_{n-1}(k+4-v_{n-1})}\nonumber\\
 &=&\exp{\left(-\frac{4(v_n-v_{n-1})}{d^2_n+d^2_{n-1}}-\frac{8}{d^2_n+d^2_{n-1}}+i4\overline{b}_n
 -\frac{(v_n-v_{n-1})^2}{2(d^2_n+d^2_{n-1})}+i\overline{b}_n(v_n-v_{n-1})\right)}\nonumber\\
 &&\quad\cdot\sum_k(k^2+2k)\exp{\left(-(k-\overline{v}^{+}_n)^2/\overline{d^2_n}
 -i(b_n-b_{n-1})(k-\overline{v}^+_n)\right)},\label{eqn:D+}
 \ea
where
$\overline{v}^+_n\equiv\frac{d^2_{n-1}v_n+d^2_n(v_{n-1}-4)}{d^2_n+d^2_{n-1}}$.
To do the summation in the above equation, we first have to rewrite
$k^2+2k$ as a function of $k-\overline{v}^+_n$:
 \ba
 k^2+2k&=&(k-\overline{v}^+_n)^2+(2\overline{v}^+_n+2)(k-\overline{v}^+_n)
 +{\overline{v}^+_n}^2+2\overline{v}^+_n\nonumber\\
&=&(k-\overline{v}^+_n)^2
+2\left(\overline{v}_n-\frac{3d^2_n-d^2_{n-1}}{d^2_n+d^2_{n-1}}\right)(k-\overline{v}^+_n)
+{\overline{v}_n}^2-2\overline{v}_n\frac{3d^2_n-d^2_{n-1}}{d^2_n+d^2_{n-1}}
+\frac{8d^2_n(d^2_n-d^2_{n-1})}{(d^2_n+d^2_{n-1})^2},\nonumber
 \ea
where $\bar{v}_n=\frac{d^2_{n-1}v_n+d^2_nv_{n-1}}{d^2_n+d^2_{n-1}}$. The sum about the square term is
 \ba
 &&\sum_k(k-\overline{v}^+_{n})^2
 \exp{\left(-(k-\overline{v}^{+}_n)^2/\overline{d^2_n}
 -i(b_n-b_{n-1})(k-\overline{v}^+_n)\right)}\nonumber\\
   &=&\sqrt\pi\sqrt{\overline{d^2_n}}\exp{\left(-(b_n-b_{n-1})^2\overline{d^2_n}/4\right)}
\left(\overline{d^2_n}/2-(b_n-b_{n-1})^2(\overline{d^2_n})^2/4\right).\nonumber
 \ea
The sum about the linear term is proportional to
 \ba
 &&\sum_k(k-\overline{v}^+_{n})
 \exp{\left(-(k-\overline{v}^{+}_n)^2/\overline{d^2_n}
 -i(b_n-b_{n-1})(k-\overline{v}^+_n)\right)}\nonumber\\
&=&-i\sqrt\pi\sqrt{\overline{d^2_n}}\exp{\left(-(b_n-b_{n-1})^2\overline{d^2_n}/4\right)}
(b_n-b_{n-1})\overline{d^2_n}/2,\nonumber
 \ea
and the contributions of the zeroth order terms are proportional to
 \be
 \sum_k\exp{\left(-(k-\overline{v}^{+}_n)^2/\overline{d^2_n}
 -i(b_n-b_{n-1})(k-\overline{v}^+_n)\right)}
 =\sqrt\pi\sqrt{\overline{d^2_n}}\exp{\left(-(b_n-b_{n-1})^2\overline{d^2_n}/4\right)}.\nonumber
 \ee
Collecting the above results, we get
 \be
 D^+_{n,n-1}=\inner{\cog{n}}{\cog{n-1}}e^{-\frac{8}{d^2_n+d^2_{n-1}}}e^{i4\overline{b}_n}
 \left((\overline{v}_n)^2-2\overline{v}_n\frac{2d^2_n}{d^2_n+d^2_{n-1}}+\frac{\overline{d^2_n}}{2}
 +P^+_{n,n-1}\right),
 \ee
where $P^+_{n,n-1}$ denotes a polynomial of $v_n-v_{n-1}$,
$b_n-b_{n-1}$ and $d_n-d_{n-1}$ without the zeroth order term. Here
we have expanded the factor
$\exp{\left(-\frac{4(v_n-v_{n-1})}{d^2_n+d^2_{n-1}}\right)}$ in
\eqnref{eqn:D+} as
$1-\frac{4(v_n-v_{n-1})}{d^2_n+d^2_{n-1}}+\cdots$. Except for the
leading term $1$, all the other terms can be conflated with
$P^+_{n,n-1}$. Analogous to the matter part, under the continuous
limit $N\longrightarrow\infty$, this $P^+_{n,n-1}$ does not
contribute to the effective action of gravity. With the experience
of computing $D^+_{n,n-1}$, it is easy to calculate $D^0_{n,n-1}$
and $D^-_{n,n-1}$ as follows:
 \ba
&&D^0_{n,n-1}=2\inner{\cog{n}}{\cog{n-1}}
 \left((\overline{v}_n)^2+\frac{\overline{d^2_n}}{2}
 +P^0_{n,n-1}\right),\\
 &&D^-_{n,n-1}=\inner{\cog{n}}{\cog{n-1}}e^{-\frac{8}{d^2_n+d^2_{n-1}}}e^{-i4\overline{b}_n}
 \left((\overline{v}_n)^2+2\overline{v}_n\frac{2d^2_n}{d^2_n+d^2_{n-1}}+\frac{\overline{d^2_n}}{2}
 +P^-_{n,n-1}\right).
 \ea
Taking the expansion
$e^{\left(-\frac{8}{d^2_n+d^2_{n-1}}\right)}=1-\frac{8}{d^2_n+d^2_{n-1}}+\mathcal{O}\left(\frac{1}{d^4}\right)$
and neglecting the higher order terms than
$\left(\frac{1}{d^2}\right)$, we can get the combination
 \ba
 &D^+_{n,n-1}-D^0_{n,n-1}+D^-_{n,n-1}\nonumber\\
 &=-4\inner{\cog{n}}{\cog{n-1}}\Big[
\Big((\overline{v}_n)^2+\frac{\overline{d^2_n}}{2}\Big)
\Big(\sin^2{(2\overline{b}_n)}\big(1-\frac{8}{d^2_n+d^2_{n-1}}\big)+\frac{4}{d^2_n+d^2_{n-1}}\Big)
 +i\sin{(4\overline{b}_n)}\overline{v}_n\big(1-\frac{8}{d^2_n+d^2_{n-1}}\big)
 +P^{\textrm{grav}}_{n,n-1}\Big],\nonumber
 \ea
and hence the matrix element
$\opelem{\cog{n}}{e^{-i\e\a\TL}}{\cog{n-1}}$ is
 \ba
\inner{\cog{n}}{\cog{n-1}}\exp\Big[-i\e\a_n\cdot3\pi{G}\Big(
\big((\overline{v}_n)^2+\frac{\overline{d^2_n}}{2}\big)
\big(\sin^2{(2\overline{b}_n)}\big(1-\frac{8}{d^2_n+d^2_{n-1}}\big)+\frac{4}{d^2_n+d^2_{n-1}}\big)&&\nonumber\\
  +i\sin{(4\overline{b}_n)}\overline{v}_n\big(1-\frac{8}{d^2_n+d^2_{n-1}}\big)\frac{2d^2_n}{d^2_n+d^2_{n-1}}
 +P^{\textrm{grav}}_{n,n-1}\Big)\Big].&&
 \ea

\end{document}